\RequirePackage[l2tabu, orthodox]{nag}
\documentclass[10pt, fleqn, final, twocolumn]{article}
\usepackage[utf8]{inputenc}
\usepackage[T1]{fontenc}
\usepackage{lmodern}
\usepackage{microtype}
\usepackage{flushend}
\usepackage{parskip}
\usepackage[dvipsnames]{xcolor}
\usepackage{xparse}
\usepackage[final]{graphicx}
\graphicspath{{./figures/}}
\usepackage[fleqn]{amsmath}
\usepackage{siunitx}
\usepackage{booktabs}
\usepackage{nccmath}
\usepackage[backend=biber, style=numeric-comp, sorting=none]{biblatex}
\renewcommand{\cite}{\supercite}
\usepackage[margin=1in]{geometry}
\usepackage{setspace}
\usepackage{amssymb}
\usepackage{fontawesome}
\usepackage{titlesec}
\usepackage[labelfont=bf, textfont=bf]{caption}
\usepackage{fancyhdr}
\usepackage{float}
\usepackage{rotating}
\usepackage{subcaption}
\usepackage{textcomp}
\usepackage{csquotes}
\usepackage{xspace}
\usepackage{relsize}
\usepackage{wasysym}
\usepackage[perpage, symbol*]{footmisc}
\usepackage[pdfusetitle]{hyperref}

\title{On the Automation, Optimization, and In-Orbit Validation of Intelligent Satellite Constellation Operations}
\author{Gregory Stock \and Juan A. Fraire \and Holger Hermanns \and Eduardo Cruz \and Alastair Isaacs \and Zhana Imbrosh}
\hypersetup{pdfsubject={Technical Report SmallSat 2021}}

\newcommand{\gomx}[1]{G{\smaller OM}X--{#1}\xspace}
\newcommand{\gomxx}{G{\smaller OM}X\xspace}
\newcommand{\gomspace}{GOM{\smaller SPACE}\xspace}
\newcommand{\POWVER}{{\smaller POWVER}\xspace}
\newcommand{\LEOPOWVER}{LEO{\smaller POWVER}\xspace}
\newcommand{\kibam}{{KiBaM}\xspace}

\newcommand{\powverHOOP}{{P{\smaller OWVER-}}{HOOP}\xspace}

\newcommand\ie{i.\,e.\xspace}

\newcommand\eg{e.\,g.\xspace}

\titleformat{\section}
{\normalfont\normalsize\bfseries}{\thesection}{0.5em}{}
\titleformat{\subsection}
{\normalfont\normalsize\bfseries\em}{\thesubsection}{0.5em}{}
\titleformat{\subsubsection}
{\normalfont\normalsize\bfseries\em}{\thesubsubsection}{0.5em}{}
\newcommand{\addstretch}[1]{\addtolength{#1}{\fill}}
\newenvironment{onepage}{%
    \newpage\flushbottom
    \addstretch{\baselineskip}
    \addstretch{\abovedisplayskip}
    \addstretch{\abovedisplayshortskip}
    \addstretch{\belowdisplayskip}
    \addstretch{\belowdisplayshortskip}
}{\newpage}

\begin{document}

\begin{onepage}
    \onecolumn
    \begin{center}
        \large % (12 point font)
        \makeatletter
        \textbf{\@title}\\
        \makeatother
        \vspace{0.5cm}
        \normalsize % (10 point font)

        {Gregory Stock}\\
        {Saarland University}\\
        {Saarland Informatics Campus E1\,3, 66123 Saarbrücken, Germany}; +49 681 302 5635\\
        {\href{mailto:stock@depend.uni-saarland.de}{stock@depend.uni-saarland.de}}\\
        \vspace{0.5cm}
        {Juan A. Fraire}\\
        {Saarland University / INRIA}\\
        {Saarland Informatics Campus E1\,3, Saarbrücken, Germany}; +54 9351 244 6010\\
        {INSA Lyon, 20 Avenue Albert Einstein, 69621 Villeurbanne CEDEX, France} \\
        {\href{mailto:juanfraire@depend.uni-saarland.de}{juanfraire@depend.uni-saarland.de}}\\
        \vspace{0.5cm}
        {Holger Hermanns}\\
        {Saarland University / Institute of Intelligent Software}\\
        {Saarland Informatics Campus E1\,3, 66123 Saarbrücken, Germany}; +49 681 302 5631\\
        {Institute of Intelligent Software, Jinzhu Plaza, No.\,221 West Huanshi Avenue, Nansha, Guangzhou, China}\\
        {\href{mailto:hermanns@depend.uni-saarland.de}{hermanns@depend.uni-saarland.de}}\\
        \vspace{0.5cm}
        {Eduardo Cruz}, Alastair Isaacs, Zhana Imbrosh \\
        {GomSpace}\\
        {GomSpace A/S, Langagervej 6, 9220 Aalborg Øst, Denmark}; +352 621 291 207\\
        {\href{mailto:ecru@gomspace.com}{ecru@gomspace.com}}

        \vspace{0.8cm}
        \textbf{ABSTRACT}\pdfbookmark[1]{Abstract}{abstract}
    \end{center}

    Recent breakthroughs in technology have led to a thriving \enquote{new space} culture in low-Earth orbit (LEO) in which performance and cost considerations dominate over resilience and reliability as mission goals.
    These advances create a manifold of opportunities for new research and business models but come with a number of striking new challenges.
    In particular, the size and weight limitations of low-Earth orbit small satellites make their successful operation rest on a fine balance between solar power infeed and the power demands of the mission payload and supporting platform technologies, buffered by on-board battery storage.
    At the same time, these satellites are being rolled out as part of ever-larger constellations and mega-constellations.
    Altogether, this induces a number of challenging computational problems related to the recurring need to make decisions about which task each satellite is to effectuate next.
    Against this background, \gomspace and Saarland University have joined forces to develop highly sophisticated software-based \emph{automated} solutions rooted in \emph{optimal algorithmic} and \emph{self-improving learning techniques}, all this \emph{validated} in modern nanosatellite networked missions operating in orbit.

    The paper introduces the \gomspace Hands-Off Operations Platform (HOOP), an automated, flexible, and scalable end-to-end satellite operation framework for commanding and monitoring subsystems, single-satellites, or constellation-class missions.
    To this, the \POWVER initiative at Saarland University has contributed state-of-the-art dynamic programming and learning techniques based on profound battery and electric power budget models.
    These models are continually kept accurate by extrapolating data from telemetry received from satellites.
    The resulting machine learning approach delivers optimal, efficient, scalable, usable, and robust flight plans, which are provisioned to the satellites with zero need for human intervention—but which are still under the full control of the mission operator.
    We report on insights gained while validating the integrated \powverHOOP approach in orbit on the dual-satellite mission \gomx{4} by \gomspace that is currently in orbit.

    \medskip
    \color[HTML]{1E76B3}
    \emph{This is an author-generated technical report of a paper published in the Small Satellite Conference 2021.\cite{conf/smallsat/StockFHCII21}}
\end{onepage}

\twocolumn

\section{INTRODUCTION}

Near-Earth satellites are being launched by the thousands; an unprecedented pace made possible by recent breakthroughs in technology accompanying a \enquote{new space} culture where cost/performance considerations dominate over resilience/reliability (\ie, emergence of COTS components and CubeSat platforms).
Although these advances create many opportunities for new research and business models, a number of striking new challenges need to be tackled in order to efficiently manage the available resources while also ensuring maximum payload utilization.
In particular, the size and weight limitations of low-Earth orbit (LEO) small satellites mean that their successful operation rests on a fine balance between solar power infeed and the power demands of the mission payload and supporting platform technologies, buffered by on-board battery storage.
This renders a non-evident, recurring, and intricate scheduling problem to be solved on the ground segment, namely the continual need to make decisions about which task the satellite is to effectuate next.
This requirement will arguably become the bottleneck for the growing trend of scaling the space segment to constellations and mega-constellations.

To this end, we contribute sophisticated software-based \emph{automated} solutions rooted in \emph{optimal} computer science techniques \emph{validated} in modern nanosatellite networked missions operating in orbit.
This paper first introduces the \gomspace Hands-Off Operations Platform (HOOP), a flexible and scalable end-to-end satellite operation framework for commanding and monitoring subsystems, single-satellites, or constellation-class missions.
By taking advantage of the vast expertise of \gomspace, new space actors can leverage flight-proven toolchains throughout the mission lifecycle while profiting from partner ground station networks without the need to invest in their own operational infrastructure.
Second, we present how HOOP is enhanced by highly efficient and accurate automated decision-making capabilities exploiting dynamic programming and learning techniques based on profound battery and electric power budget models, developed at Saarland University as part of the \POWVER initiative.
The models are continually kept accurate by extrapolating data from telemetry received from satellites.
The resulting machine learning approach delivers optimal, efficient, scalable, usable, and robust flight plans, which are provisioned to the satellites with zero need for human intervention—but which are still under the full control of the mission operator.
Third, we report on the application of the \powverHOOP approach to \gomx{4}, the dual-satellite mission by \gomspace that is currently in orbit.
Over a period of more than a month, a series of in-orbit experiments have been carried out with the 6U CubeSats, covering Earth observation, air traffic surveillance, as well as inter-satellite linking capabilities.
In these experiments, the integrated \powverHOOP toolchain has shown its unique strength, namely to operate a mission without human intervention while persistently delivering maximum return from its observation payloads and ensuring the most efficient and safe utilization of constrained on-board battery resources.
We make these findings concrete by reporting details of a 48-hour period selected from the masses of recorded experimental results.

This pioneering work evidences that humans can define and supervise high-level objectives of the mission while relying on machine learning approaches to finally unblock the future of space operations.

\section{CONTEXT}

\subsection{\texorpdfstring{\gomspace and Their Mission}{GomSpace and Their Mission}}

Since the foundation of the company in 2007, \gomspace has become a leading manufacturer and supplier of CubeSats and small satellite solutions for customers in academic, government, and commercial markets.
The key strengths of the company include systems integration, CubeSat platforms, advanced miniaturized radio technology, and satellite operations. 
The \gomspace headquarters are located in Aalborg, Denmark.
The company also has a propulsion technology center in Uppsala, Sweden, and a satellite operations center in Esch-sur-Alzette, Luxembourg.
The company currently employs more than 150 people and provides services to customers in more than 60 nations.

\gomspace has a track record of successful missions in space.
This is exemplified by the \gomxx{} series of satellites, all of which were built and operated by \gomspace.
\gomx{1}, a 2U satellite launched in November 2013, successfully demonstrated for the first time the reception of ADS-B signals from aircraft by an orbiting satellite.
The satellite remains in orbit.
\gomx{3}, launched in 2015, demonstrated attitude control, downlinking of data, and SATCOM spot-beam characterization.
The satellite successfully completed its nominal mission and re-entered the atmosphere after one year. 
That mission was followed by the \gomx{4} mission, a pair of two 6U CubeSats.
This mission demonstrated the ability of CubeSats to act in coordination through inter-satellite communication.
Payloads of the satellites are used for surveillance and monitoring of Arctic regions.

These projects are delivered based on \gomspace's strong in-house portfolio of established products and rich capabilities.
Currently, \gomspace is developing the Juventas nanosatellite that will form part of ESA's Hera mission to the Didymos binary asteroid system.\cite{conf/smallsat/GoldbergKRHTP19}
The mission will provide valuable scientific data from the asteroid system, including radar and radio science observations of the binary system.

With more than 13 years of experience in the market and a track record of multiple successful missions accomplished, \gomspace has developed profound knowledge and competencies within radio technology, CubeSat platforms, project management and innovation.
Starting early as a pioneer in the market, \gomspace has now become a market leader in the commercialization of nanosatellites and new space technology.

The numerous lessons learned by \gomspace during past satellite projects and operations materialize into HOOP. 
Described in detail below, HOOP is a platform enabling automated satellite operation at low cost that is individually adapted to the specific mission requirements.
As evidence of HOOP's flexibility, this paper presents its seamless integration with the \LEOPOWVER toolchain.

\subsection{\texorpdfstring{\POWVER at Saarland University}{Powver at Saarland University}}

Since 2013, Saarland University is performing scientific research on the operation of nanosatellites, seeded in the EU FP7 project SENSATION\footnote{\url{https://cordis.europa.eu/project/id/318490}}, where \gomspace acted as an industrial partner.
The unique expertise of the Saarland University researchers lies in the application of formal methods to perform automatic and resource-optimal task scheduling of satellites and satellite constellations.
Early work developed the scientific grounds along the \gomx{1}\cite{DBLP:journals/lites/HermannsKN17} and \gomx{3} missions, including in-orbit demonstrations\cite{journals/actaastro/NiesSKHBG18, DBLP:journals/fac/BisgaardGHKNS19}.
In these contexts, it had become apparent that there is massive room for improvement by properly modeling and analyzing the satellite's battery, operational constraints, and orbital environment.
The research intensified as a consequence of a multi-million dollar award, the ERC Advanced Grant \POWVER\footnote{\url{https://cordis.europa.eu/project/id/695614}}, that was awarded in 2016 to Holger Hermanns by the European Research Council to foster the research.
Another (albeit smaller) grant, the ERC Proof of Concept Grant \LEOPOWVER\footnote{\url{https://cordis.europa.eu/project/id/966770}}, is nowadays the focus point of all application and commercialization activities of what has been developed successfully over the years: \emph{orbit-proof software enabling the continuous, fully automated, energy-optimal, and profit-maximizing dynamic operation of LEO satellites and satellite constellations}.

At the core of the \LEOPOWVER software is a unique collection of highly realistic battery models that enable efficient forecasting of battery health and battery depletion risk with unprecedented accuracy.
This is paired with ultra-efficient optimization and machine learning techniques that are tailored to the LEO context, including dedicated support for telemetry processing and contact plan design for satellite constellations.\cite{DBLP:journals/cm/FraireF15}
Indeed, communication transmitters and transponders are among the most power-demanding subsystems of any modern spacecraft.
This phenomenon is exacerbated in networked space constellations supported by one or more Inter-Satellite Link (ISL) interfaces.
As a result, a notorious bottleneck is provoked by power-hungry networking tasks that need to be powered by constrained batteries and solar power infeed.
This asks for a very careful time-evolving and data-driven scheduling of the communication resources.

The methodologies delivered by \LEOPOWVER are targeted at the core of this problem, which needs to be perpetually solved during the mission lifetime.
Thus, the role of the operator is reduced to the most important aspects of the mission: defining the goal and the conditions to achieve it, leaving computer science algorithms to ensure the optimal and secure control of the space system.
\LEOPOWVER harvests very advanced algorithmic and learning approaches, which are already unlocking the optimal battery-aware operations of future cross-linked satellite constellations.\cite{DBLP:journals/tgcn/FraireNGHBB20, DBLP:journals/ijscn/FraireGHNBB21}

\begin{figure*}[t]
    \centering%
    \includegraphics{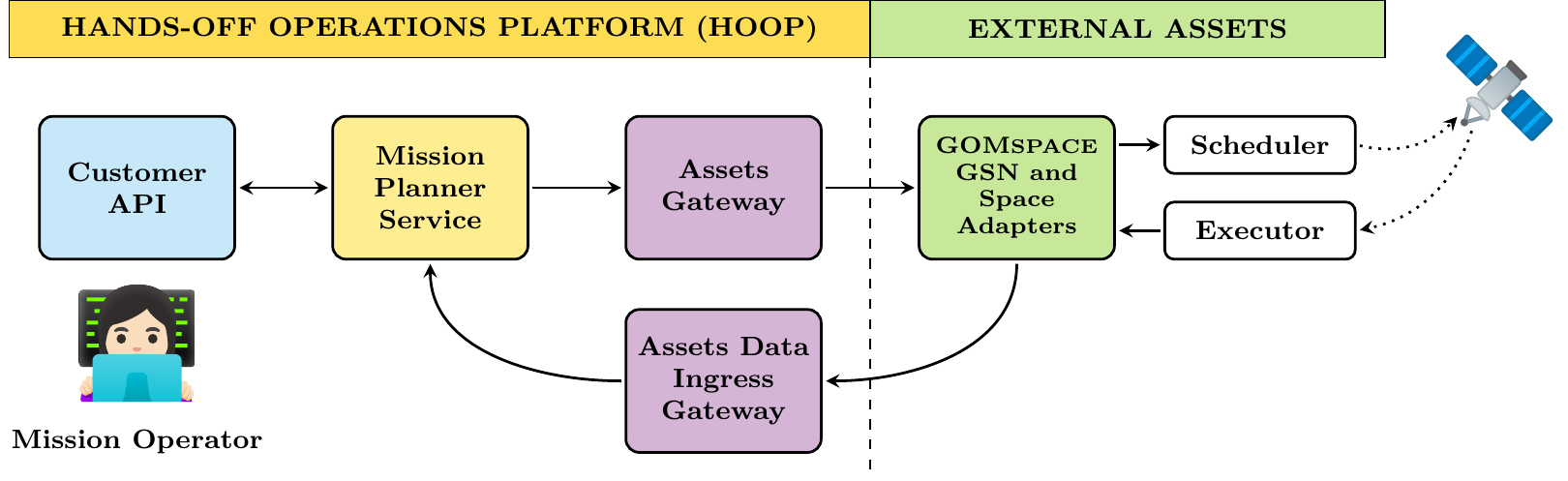}
    \vspace{4pt}
    \caption{Overview of the \gomspace Hands-off Operations Platform (HOOP).}%
    \label{fig:hoop-overview}%
\end{figure*}

The spirit of \LEOPOWVER is to make this unique combination of technologies ready for take-up by embracing start-ups, academics, and business customers with the intention to create a prospering user community.
Thus, favorable conditions will be enabled via a base version of the software to be released under open-source licensing in the near future.\footnote{Check \url{https://leopowver.space/} for the latest updates.}
To this end, the \LEOPOWVER orchestration toolchain is designed to provide highly flexible support for integration into arbitrarily complex operations workflows via well-defined telemetry and commanding interfaces across the entire spectrum of orbit applications. 

In this paper, we describe how this flexibility is exploited for a straightforward and successful integration of \LEOPOWVER with the \gomspace HOOP operations platform to then validate its applicability in the state-of-the-art \gomx{4} nanosatellite mission currently in orbit.

\section{HANDS-OFF OPERATIONS PLATFORM}

The \gomspace Hands-off Operations Platform (HOOP) is a satellite operations platform built for automation, scalability, and flexibility.
The HOOP platform was developed by \gomspace specifically to handle operations for constellations of CubeSats, up to constellations with thousands of satellites. 
Since 2018, the platform has been under development with support from the Luxembourg Space Agency (LSA) and European Space Agency (ESA).

Unlike traditional operations centers, which carry out operations manually with large numbers of satellite operators, operations centers for constellations must automate much of the nominal operations.
The HOOP system has been designed to support and manage this degree of operations, and thereby allow operations to scale to support hundreds or thousands of satellites.

In this way, HOOP streamlines routine operations, allowing operators to focus on troubleshooting and improving the mission.
This reduces the manual effort required to monitor and maintain each satellite, and also means that routine activities, such as payload data capture and download, can be completely automated and require no manual input at all.

HOOP provides a set of distinguishing features for the management and autonomous operation of satellites, from the first satellite to a global constellation.
A schematic overview displaying all components that HOOP consists of is shown in \autoref{fig:hoop-overview}.

\paragraph{Configuration Management.}
A difficult challenge in satellite operations is to track and know the satellite configuration at all times, even when the satellite is out of visibility of a ground station.
HOOP addresses this challenge by providing a set of configuration management tools.
The last-known satellite configuration is stored in a database and visible to operators through a user interface.
All changes planned through HOOP are recorded, and all downlinked telemetry is monitored for any discrepancies with the expected configuration.
Operators are alerted if any unexpected configuration changes are detected.
The database used by HOOP allows the configuration of multiple satellites to be tracked.
When an issue arises with a satellite, an operator can call up the relevant satellite from the database and review the last known configuration of all on-board parameters.

\paragraph{Mission Planning.}
The operations software handles both manual and automated mission planning.
Operators can define routine procedures that are scheduled according to specified rules (based on events or time).
They can also plan manual procedures as required, expressed in a high-level yet flexible procedure language.
A plan resolver checks all operations plans, generated either manually or automatically, and verifies that there are no conflicts present.
If conflicts are found, the resolver attempts to resolve them.
If no solution can be found, it will reject the operations plan and send an alert to the operator.

\paragraph{Contact Scheduling.}
After procedures are validated, they are scheduled for upload and execution on the satellite.
HOOP can handle multiple ground stations and will schedule the upload on the next available pass for the satellite over any of the available ground stations.
Once the pass begins, the scheduled commands are executed, and prepared files are uploaded.
The command status is always visible to the operator, so they are aware if the commands are planned, have been executed, or failed.
If a command fails to execute on the satellite or a file upload fails, HOOP can be configured to either abandon the effort or retry.

\paragraph{Payload Data Download.}
The download of collected payload data from the satellite is scheduled autonomously by HOOP.
The best opportunities for downloading the data are calculated, considering any conflicts and resource availability (\ie, available data storage).
After new data is downloaded, HOOP transfers it to the end user.

\paragraph{Telemetry Handling.}
Telemetry received from the satellite is stored in a telemetry database.
Operators can visualize real-time and historical telemetry data using HOOP's telemetry dashboard.
This dashboard offers a flexible and easily configurable view of the data.
Operators can select the data they want to see, and can quickly request data from different time periods and with different granularity.
This telemetry data is also visible to users through the customer API (application programming interface).

\paragraph{Alarms.}
Operators can define rules for raising alarms.
HOOP constantly monitors the incoming telemetry data and will raise alarms when the telemetry meets conditions defined in those rules.
Operators are informed through the platform and through notifications, which can be sent via email or to mobile devices.
An alarm console records all active alarms.
Operators can also define rules for automatically responding to specific alarms.
This feature will only be used for alarms that are well understood and have a known resolution.
All automatically taken commands are logged and available for review by an operator.

\enlargethispage{.1\baselineskip}

\paragraph{Flight Dynamics.}
HOOP uses Orekit\cite{OREKIT} for internal flight dynamics calculations, including the calculation of ground station passes and orbital events.
HOOP is also compatible with tools developed in Orekit or GMAT\cite{GMAT} for maneuver planning, \eg, for formation deployment, station keeping, or collision avoidance.

\paragraph{Simulation.}
HOOP incorporates a number of simulation tools for mission planning.
This includes data, thermal, and link simulators.
These tools analyze the current satellite state, based on the telemetry and known configuration, and generate forecasts of future performance.
These tools may also be used to validate requested procedures and flight plans to ensure that no constraints (\eg, thermal or data limits) are breached.

\paragraph{Customer API.}
Through the customer API, the mission operator has full control of the mission.
There are two main API endpoints: the \emph{mission planning} interface and the \emph{telemetry \& payload data} interface.
The former can be used to control the various payloads on the satellite via external planning software, while the latter allows the customer to oversee the mission by providing access to downloaded telemetry data as well as collected payload data, \eg, images taken by a camera payload.
Operation requests are represented as generic lists of time periods indicating when each payload module should be activated.
This allows the mission operator to focus on his interests (\eg, with respect to certain areas of interest where Earth observations should be acquired) and compile a matching schedule without the need to think about required attitude control or preheat constraints.
This step is handled by the mission planner.
The plan resolver converts the scheduled payload activation intervals to a flight plan that can be uploaded to the satellite, thereby adding the necessary slewing or preheating maneuvers.
After the feasibility of the resulting plan is ensured and potential conflicts are resolved, the planner finds the next ground station pass in which the plan can be uploaded to the satellite and triggers an upload request to the \gomspace ground station network (GSN).

The mission planning API and the telemetry \& payload data API are the natural interfaces for integration with the \LEOPOWVER toolchain described below.

\section{\texorpdfstring{\LEOPOWVER INTELLIGENCE}{LEOPOWVER INTELLIGENCE}}

\enlargethispage{.1\baselineskip}

\begin{figure*}[t]
    \centering%
    \includegraphics[width=.95\textwidth]{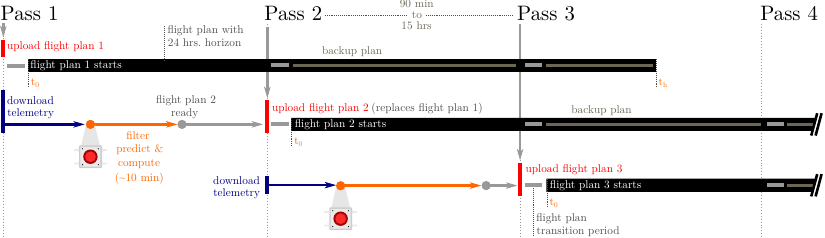}
    \caption{Receding-horizon scheduling along different passes over the ground station. The computed flight plan extends well beyond the needed time horizon to serve as a backup plan in case that the next update fails.}
    \label{fig:receding-horizon}
\end{figure*}

This section presents the central features and components of the \LEOPOWVER software infrastructure for safe, fully automated, energy-efficient, and intelligent operation of LEO satellites and satellite constellations. 

At its core, \LEOPOWVER takes over the task of continuously maximizing payload utilization while eliminating the risk of overstraining the power budget at any moment.
The approach is flexible in the way that it can express the intentions of spacecraft engineers and mission operators with respect to the finer optimization goals.

While running, \LEOPOWVER continually triggers a sophisticated decision-making procedure to deliver optimal schedules ready to be uploaded before each ground station pass.
Although the tool is able to work autonomously, a human operator can be kept in the loop to oversee and safeguard the entire operation.
An \textit{orchestrator} acts as the central coordinator of the automated process within the toolchain involving mission models, deep battery models, and a scheduler engine as described below.

The overall process is a resourceful variant of receding-horizon scheduling, a well-known approach\cite{DBLP:journals/msom/ChandHS02} to perpetuated finite-horizon scheduling.
Our instantiation of the receding-horizon principle is visualized in \autoref{fig:receding-horizon}.
In a nutshell, the tool can be configured to calculate schedules over some predefined time interval~$I$ in the order of one or two days.
These schedules are continually re-computed based on the latest available information, \ie, telemetry.
Whenever the satellite passes over the ground station, the most recent schedule is offered for upload.
Owing to the LEO context, the time span between any two consecutive passes lies somewhere between 90~minutes and 15~hours.

\begin{figure*}[t]
    \centering%
    \includegraphics[width=\linewidth]{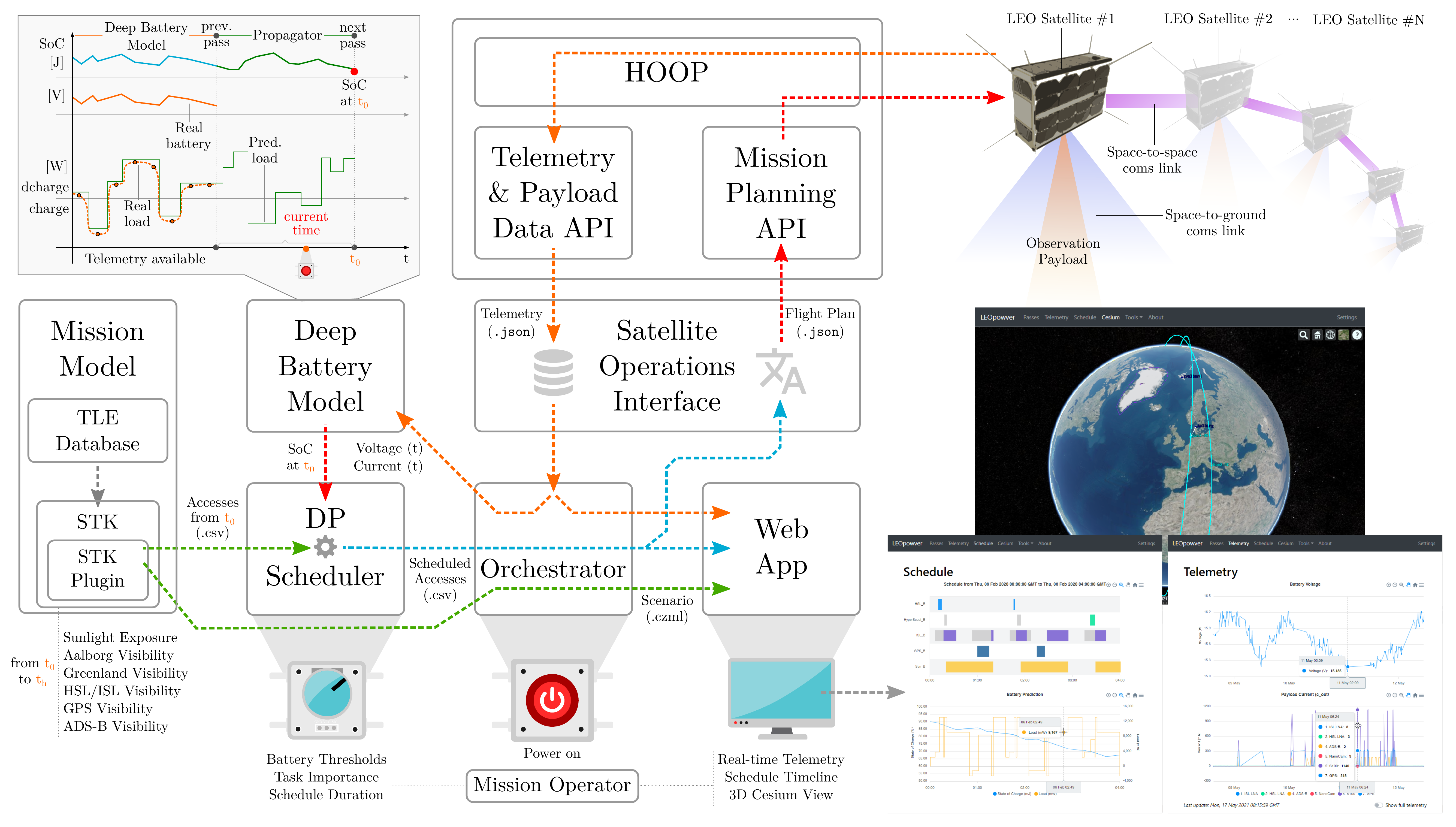}
    \caption{The \LEOPOWVER toolchain interfacing with HOOP for the operation of LEO constellations.}
    \label{fig:architecture}
\end{figure*}

\autoref{fig:architecture} shows a schematic overview of all components the \LEOPOWVER toolchain consists of.

\paragraph{Orchestrator.}
The Orchestrator takes care of coordinating the availability of up-to-date information within the toolchain.
In particular, it interfaces with the satellite operator's API in order to receive the latest telemetry of the satellite.
The telemetry, in particular logs of voltage and current measurements of the satellite's battery, is fed to the Deep Battery model.
And the flight plans provided by the DP module are transferred into the required format and delivered to the satellite operator's API, after being checked for plausibility and feasibility.

\paragraph{DP Scheduling Engine.}
The dynamic programming (DP) scheduling engine is the central algorithmic component of \LEOPOWVER.
It receives the access windows regarding the different payloads on board, together with the background load and state of charge estimates regarding the battery at $t_0$ (the beginning of the scheduling interval $I$).
The DP delivers a set of tasks to schedule that is optimal with respect to the stated objectives, ensuring that the battery stays above a minimum state of charge.
A natural (yet simple) example for such an objective would be to assign a certain reward to each executed task and then to maximize the total reward accumulated.
Notably, the objective is highly customizable and easy to adapt to the mission operator's needs.

Dynamic programming\cite{Bellman57} proved to be a very efficient way to solve these kinds of optimization problems.
Partial solutions that potentially lead to optimal solutions are stored in an efficient data structure called the \emph{DP table}.
The existing entries are then used to further explore the options and to gradually fill the table.
Efficiency directly correlates with the size of the DP table, \ie, the number of entries.
Therefore, highly sophisticated antichain-based pruning methods are implemented to remove unnecessary entries and exclude them from further analysis.\cite{DBLP:journals/tcad/StockFMHBC20}
For maximum performance, this part of the toolchain is implemented in Rust\footnote{\url{https://www.rust-lang.org/}}.

\paragraph{Deep Battery Model.}
One of the innovative contributions to the field of power-aware scheduling is the integration of the Kinetic Battery Model (KiBaM) in the workflow of cost-optimal task scheduling.
The KiBaM stores its charge in two parts, namely the available charge $a(t)$ and the bound charge $b(t)$.
When a load $\ell (t)$ is applied on the battery, only the available charge is consumed instantly, while the bound charge is slowly converting into available charge via diffusion, as such representing chemically bound energy inside the battery.
Diffusion can also happen in the other direction, depending on the amount of both types of charges.
One can think of the KiBaM as two wells holding fluid, interconnected by a small pipe.
A non-negative diffusion rate~$v$ controls the diffusion speed.
The two-dimensional state of charge (SoC) of the KiBaM is mathematically evolving according to the two coupled differential equations $\dot{a}(t) = -\ell(t) + v \cdot \big(b(t) - a(t)\big)$ and $\dot{b}(t) = v \cdot \big(a(t) - b(t)\big)$, assuming that both wells have the same capacity.
Unlike linear battery models, the KiBaM captures a number of non-linear effects of real batteries, like the recovery effect and the rate-capacity effect.\cite{DBLP:journals/iee/JongerdenH09}

Within \LEOPOWVER, the battery model is able to learn and adapt based on the latest in-orbit measurements, as displayed in the upper left of \autoref{fig:architecture}.
To this end, the SoC of the battery at the beginning of the scheduling period is obtained.
The Deep Battery model exploits the available telemetry of the satellite to properly estimate the real load the battery was feeding energy to and the battery voltage that is used to learn the estimated SoC during the period.
While earlier work has used a relatively simple Kalman filter,\cite{DBLP:journals/fac/BisgaardGHKNS19, DBLP:journals/tcad/StockFMHBC20} we are now using more advanced intelligence for this purpose.
Both battery voltage and current telemetry are sampled on the satellite, which together can be used to determine the output and input load of the battery at least once every two minutes (apart from gaps in the downlinked data due to other storage and transmission priorities).
By means of current integration (a.\,k.\,a. Coulomb counting\cite{DBLP:journals/lites/HermannsKN17}), the evolution of the stored energy can be properly approximated, and the actual SoC at the end of the telemetry period can be determined while learning adjustments to the battery model parameters via advanced learning techniques.
Next, the propagator operates for the period when no telemetry is available and propagates the battery SoC using the predicted (scheduled) load and the kinetic battery model.
At the end of the process, the SoC at time $t_0$ is made available to the DP scheduler.

\paragraph{Mission Model.}
The \LEOPOWVER intelligence relies on accurate mission abstractions for which STK~\cite{STK} is leveraged.
To this end, a dedicated STK plugin interfaces with STK core functions to access state-of-the-art orbital propagators, advanced sensor modeling, and access calculations.

The module can interface with a database to obtain the latest orbital parameters of the satellites, as well as precise ground station location, altitude, and elevation profiles.
In this, the Two-Line Element (TLE) format, which encodes in two lines of text all the necessary Keplerian parameters defining the orbit of the satellites, is leveraged.
Afterwards, STK's built-in Simplified General Perturbations~4 (SGP4)\cite{vallado2006revisiting} algorithm is used to propagate the trajectories of the satellites into the future.
SGP4 is based on accurate analytical and numerical methods considering perturbations caused by the Earth's shape, drag, radiation, and gravitational effects from other bodies such as the Sun and Moon.
If needed, the \LEOPOWVER toolchain is prepared to work with more precise propagators such as HPOP\cite{refaat2018high}.

The resulting trajectories are enriched by sensors mimicking the antenna and the coverage of each payload in the evaluated mission (missions with more complex RF payloads can also profit from the Communications module of STK).
Next, we capture the contact windows between ISL sensors among flight segments, as well as space to ground segment contacts.
To this end, the rich visibility constraints interface provided by STK is leveraged, \eg, ground station elevation angle profiles, maximum link distance, and sun angle exclusion.
On the ground part of the model, the territories of interest are modeled by polygons, and the ground stations of the mission are placed at their actual site.
We also care about sunlight access for each of the satellites in order to model the time episodes in which the on-board batteries are charged.
Sunlight exposure is computed internally in STK using accurate planetary dynamics and exported to the DP engine in the form of simple tables indicating the start and end time of each exposure episode.

Based on this STK scenario, the plugin is able to compute relevant accesses, which are delivered to the DP scheduler in \texttt{.csv} format.
When needed, obtained windows can be partitioned to allow the DP module to take granular scheduling decisions.
Furthermore, the STK script also exports the resulting scenario as a \texttt{.czml} file.
This file is then fed to a Web App that uses the CesiumJS library\footnote{\url{https://cesium.com/platform/cesiumjs/}} to visualize an interactive 3D view of the scenario accessible via any modern web browser.
The whole scenario can be intuitively explored by manipulating the simulation time using the playback and time control features.
All of the described processes are fully automated and require no human intervention whatsoever.

\paragraph{Satellite Operation Interface.}
This interface bridges the gap between real satellites and the \LEOPOWVER approach to dynamic programming and learning.
It serves two main purposes:
First, this module collects telemetry measurements and stores them in an internal database where the Deep Battery model can access the data.
And second, it sends computed flight plans to the satellite operator's API, thereby converting the abstract schedule representation into the desired output format that can be uploaded to the satellite.
Since these are the only two interfaces to external systems, an integration of the \LEOPOWVER tool with existing satellite operation frameworks like HOOP is fairly easy.

\section{\texorpdfstring{\gomx{4} MISSION OVERVIEW}{GOMX-4 MISSION OVERVIEW}}

Before providing details regarding the in-orbit test campaign carried out with the combination of HOOP and \LEOPOWVER, we provide an overview of the \gomx{4} mission and its objectives. 

The \gomx{4} program is a research and development mission led by \gomspace in partnership with the Danish Defense Acquisition and Logistics Organization (DALO), the Technical University of Denmark, and the European Space Agency.
The mission consists of two 6U CubeSats, \gomx{4A} and \gomx{4B}, both of which carry a number of payloads.
These payloads demonstrate key enabling technologies for future nanosatellite constellations, namely orbit maintenance, inter-satellite communication, high-speed downlinking, and advanced remote sensing.\cite{conf/smallsat/LeonKW18}

\begin{figure*}[t]
    \centering%
    \includegraphics[width=\linewidth, trim={0 38 0 13}, clip]{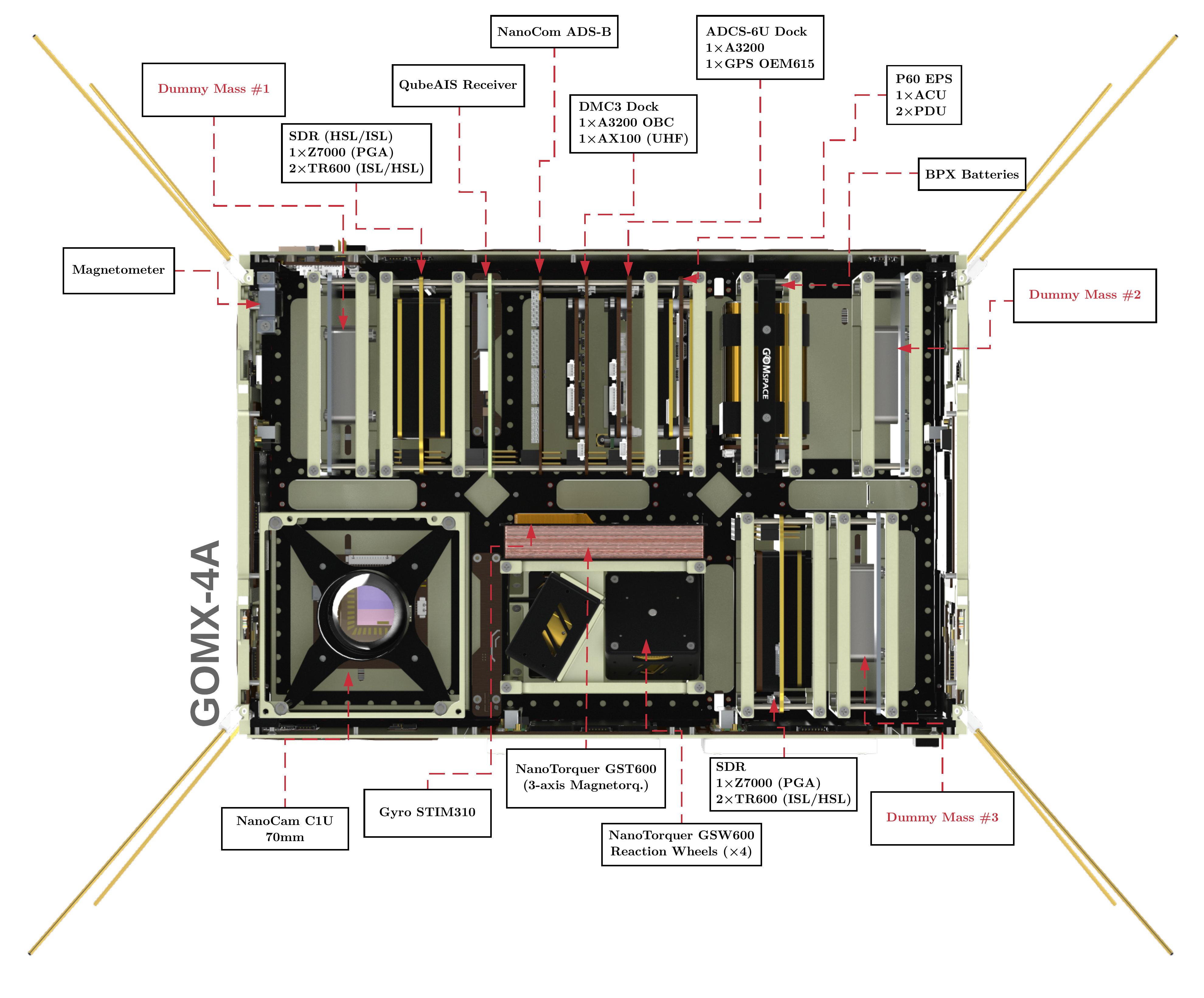}
    \caption{Internal layout of \gomx{4A}.}%
    \label{fig:gomx4a-internal-layout}%
\end{figure*}

The two satellites were launched into a \SI{500}{\km} altitude Sun-Synchronous Orbit (SSO) on February~2, 2018, from the Jiuquan Satellite Launch Center in China.
Following six weeks of commissioning activities, the satellites entered a six-month technology demonstration phase, during which the payload and operations concept was demonstrated.
The satellites have an expected lifetime of 3 to 5 years.

Inter-satellite communication is considered a key enabling technology for future nanosatellite constellations.
The \gomx{4} mission demonstrated the functioning of an S-band link between the \mbox{\gomx{4A}} and \gomx{4B} satellites.
To allow the link to be established, an inter-satellite separation of less than \SI{4500}{\km} must be maintained.
To ensure this, the \gomx{4B} satellite is fitted with a cold gas propulsion system.
No propulsion system is present on the \gomx{4A} satellite.
The inter-satellite link was demonstrated during the technology demonstration phase.

The concrete application context of the \gomx{4A} mission is the surveillance and collection of remote sensing data of the Arctic region, especially the Greenland territory.
For this purpose, the satellite is fitted with an imaging device and with an ADS-B receiver for aircraft tracking.
The satellite is also known under the name Ulloriaq, the Greenlandic word for \enquote{star}.

Both satellites are commanded from a ground station located in Aalborg, Denmark (\ang{57;01;22}N, \ang{9;58;41}E).
Two antennas are available at this ground station, one in UHF and one in S-band.
The UHF antenna is used for robust low-data rate tasks such as telecommands and the reception of telemetry.
The S-band antenna is used to download payload data.
Due to the orbit properties of the SSO, each satellite is visible to the ground station five to six times per day.
These contact opportunities are clustered, typically in the early morning and late afternoon each day.

\subsection{Platform}

The platform of the \gomx{4} satellite (see \autoref{fig:gomx4a-internal-layout}) maintains the stability and functioning of the satellite.
The satellite follows the standard \gomspace 6U architecture, \ie, a decentralized architecture where the failure of any one node does not impact the functioning of any other.

\begin{itemize}
    \item \textbf{EPS.}
          The Electrical Power System consists of a NanoPower P60.
          This system manages input power from the solar panels fixed to the outside of the satellite and supplies power to the other components on board.
          Power is stored using a BPX battery subsystem.
    \item \textbf{Solar Panels.}
          \gomx{4} uses solar panels to obtain electrical power.
          These panels are positioned on one side of the satellite, meaning the satellite must be orientated correctly to achieve maximum charging.
    \item \textbf{ADCS.}
          The ADCS subsystem determines and controls the satellite attitude.
          The system on \gomx{4} uses sun sensors, magnetometers, and magnetorquers to perform this function.
    \item \textbf{OBC.}
          The On-Board Computer (OBC) provides processing and data storage capabilities.
          Flight plans activated on the OBC may instruct other components on the satellite according to scheduled commands.
    \item \textbf{UHF Link.}
          The UHF Link enables communication between \gomx{4} and the ground station at Aalborg.
          Telemetry and telecommands (\ie, flight plans) are transferred via this link.
    \item \textbf{High Speed Link.}
          The High Speed Link (HSL) uses an S-band radio link to provide a fast data connection to the ground station at Aalborg, used to transfer acquired payload data to the ground segment.
\end{itemize}

\subsection{Payloads}
The payload of the \gomx{4A} satellite is composed of a number of sensors designed for communication and surveillance.

\begin{figure*}
    \centering%
    \includegraphics{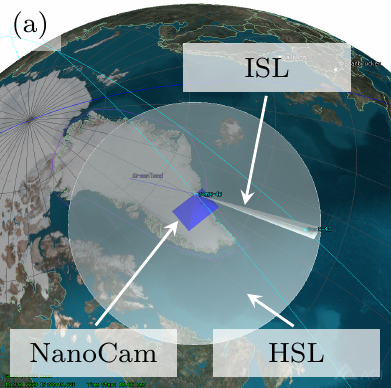}\hfil
    \includegraphics{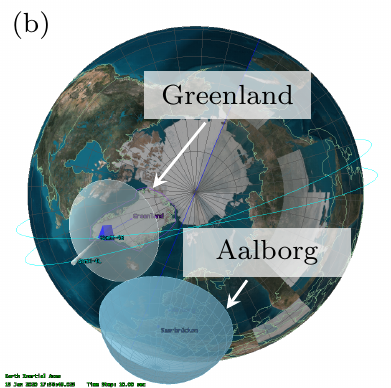}\hfil
    \includegraphics{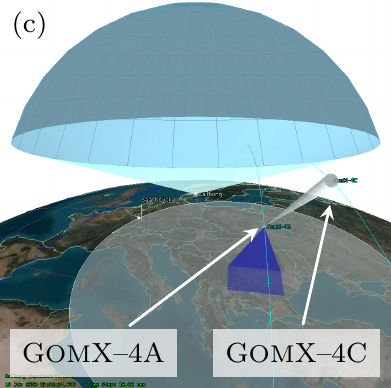}\hfil
    \includegraphics{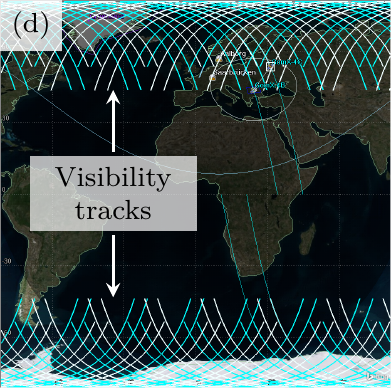}
    \caption{Different \gomspace mission scenario situations visualized in STK.}%
    \label{fig:stk-screenshots}%
\end{figure*}

\begin{itemize}
    \item \textbf{NanoCam Camera.}
          The NanoCam C1U Camera is an RGB imaging device that is used to observe the Greenland territory.
    \item \textbf{Inter-Satellite Link.}
          The Inter-Satellite Link (ISL) enables data transfer between the \mbox{\gomx{4}} twins and other LEO satellites close enough (within \SI{4500}{\km}) that are able to communicate in S-band.
    \item \textbf{Global Positioning System.}
          The new Novatel OEM719 payload is an additional Global Positioning System (GPS) receiver that will eventually replace the Novatel OEM615 also on-board the satellite.
          It enables \gomx{4A} to receive GPS satellite location signals to determine its precise local position.
    \item \textbf{ADS-B Receiver.}
          The Automatic Dependent Surveillance--Broadcast is a surveillance technology allowing the tracking of the position, velocity, and other sensor data broadcast by aircraft.
          It is used by \gomx{4A} to track aircraft over the Atlantic where ground-based reception is difficult.
\end{itemize}

UHF is enabled whenever Aalborg is in line-of-sight, so it can be considered as a recurring background load on the battery, similar to sunlight exposure (but with opposite effects on the battery).
ISL transceivers are installed on board for gaining experience with data transfer between satellites, whenever possible over the poles.
The restriction to over-the-pole episodes is rooted in the necessity to comply with international radio frequency regulations regarding the use of \mbox{S-band} over inhabited regions.

\section{EXPERIMENTAL RESULTS}

This section presents the empirical evaluation results of a thorough test campaign where \gomx{4A} has been operated by the \powverHOOP toolchain.
During a period spanning from April 12 to May 14, a variety of different in-depth experiments were carried out in orbit.

\subsection{Scenario Set-Up}
The concrete mission scenario we focus on in detail consists of the \gomx{4A} satellite and an imaginary satellite \gomx{4C} acting as a receiver/transmitter of data in the constellation.
The latter is placed in orbit in such a way that the resulting communication opportunities are non-trivial to plan for.
For this, the STK plugin computes the corresponding TLE of the imaginary \gomx{4C} satellite, based on the actual data for \gomx{4A} with RAAN angle and mean anomaly shifted by \ang{10}.
The RAAN angle indicates the angle of the orbital plane in the equator, while the mean anomaly stands for the position of the satellite along the trajectory in the orbit.

ISL tasks are allowed to occur only when the ISL sensor on \gomx{4A} can point to \gomx{4C} on a distance no larger than \SI{1300}{\km}.
As illustrated in \autoref{fig:stk-screenshots}, the resulting configuration renders two satellites that separate maximally at the equator and become aligned in an along-track configuration as they come close over the North and South poles.
Exactly in this condition, the inter-satellite distance is minimal, and the inter-satellite link antennas become aligned.
As a result, ISL transmission between \gomx{4A} and \gomx{4C} is only possible during over-the-pole episodes.
Thereby, potential interference with ground radio-frequency services operating on S-band in populated regions is avoided.
Depending on the small orbital variations, these episodes last around 20 minutes.
Since the power demand of the ISL payload is considerable, we allow for intermittent utilization of the resource.
To account for this, the ISL windows are partitioned into three smaller chunks of 7 minutes each.

Flight plans are usually uploaded via the UHF link during passes over the ground station in Aalborg.
Since the process of swapping the flight plan needs exclusive access to the satellite, and to make the scenario more interesting, we introduce two different ground stations to the scenario that are used to downlink collected data via HSL: Teófilo Tabanera Space Center in Córdoba, Argentina (\ang{31;31;30}S, \ang{64;27;46}W) and Svalbard Satellittstasjon in Svalbard, Norway (\ang{78;13;47}N, \ang{15;24;28}E).
According to \gomx{4A} specifications, a conic HSL sensor is configured with \SI{2000}{\km} maximum range and \ang{65} of half-angle pointing in the nadir direction (\ie, to ground) while the HSL downlink antennas of the two ground stations in Córdoba and Svalbard are modeled each by a sensor with a half-angle of \ang{70} and maximum range of \SI{2000}{\km}.

\begin{figure*}
    \centering%
    \includegraphics{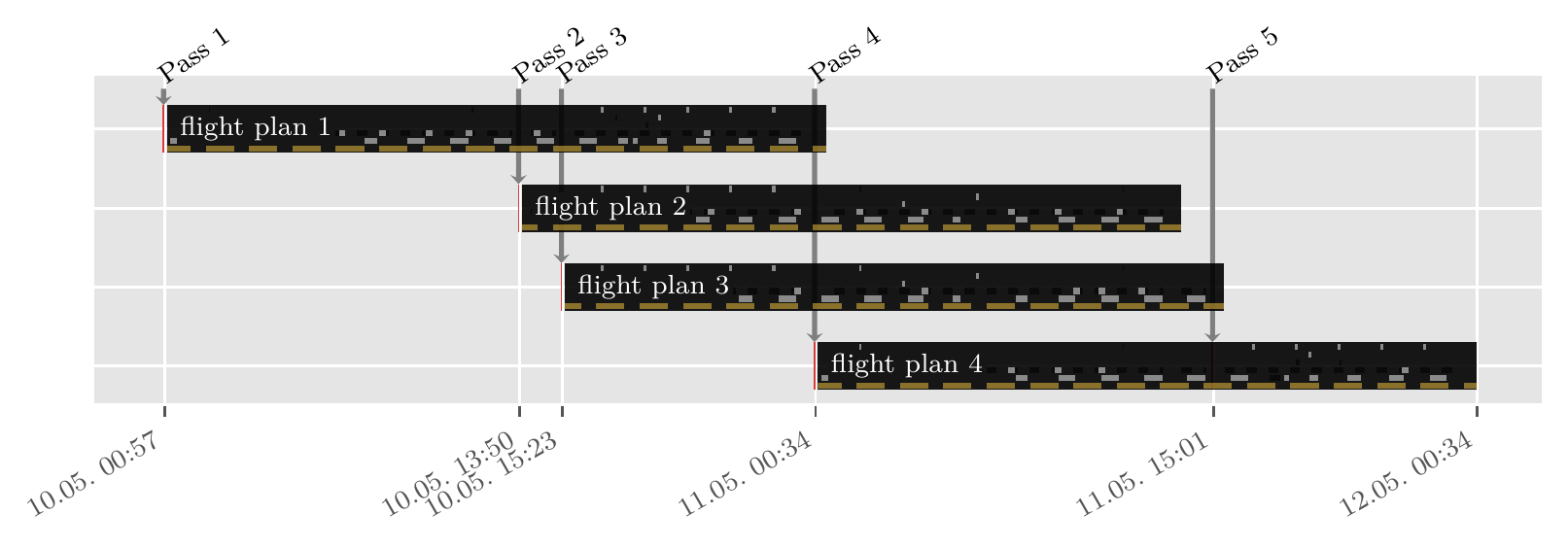}
    \caption{Experiment overview with the four uploaded flight plans.}%
    \label{fig:overview-four-plans}%
\end{figure*}

The NanoCam is modeled by a rectangular sensor pointing in nadir direction with \ang{16} and \ang{23} of across and along angles, respectively.
Visibility conditions between the NanoCam and the Greenland territory---that is modeled by a polygon---are stored as potential camera tasks.

Finally, \gomx{4A} is configured to use its \mbox{ADS-B} receiver to track aircraft only when orbiting over the Atlantic Ocean, where coverage with terrestrial antennas is difficult.
While the Atlantic is again modeled by a polygon, the ADS-B receiver has no attitude constraints of interest and uses a simple line of sight constraint.

\begin{table}
    \centering%
    \newcommand{\statusWarn}{\textcolor{Goldenrod}{\footnotesize\faExclamationTriangle}}%
    \newcommand{\statusOk}{\textcolor{Green}{\small\faCheckSquare}}%
    \captionof{table}{Experiment Summary.}
    \label{tab:experiment-summary}
    \begin{tabular}{ccl}
        \toprule
        Start of Schedule       & Max El.     & Status                    \\
        \midrule
        May 10, 00:57           & \ang{85.85} & \statusOk{} Executed      \\
        May 10, 13:50           & \ang{36.22} & \statusOk{} Executed      \\
        May 10, 15:23           & \ang{28.42} & \statusOk{} Executed      \\
        \midrule
        May 11, 00:34           & \ang{53.59} & \statusOk{} Executed      \\
        May 11, 15:01           & \ang{44.67} & \statusWarn{} Backup Plan \\
        \bottomrule
    \end{tabular}
    \vspace{1ex}

    \statusOk{} -- flight plan successfully uploaded \\
    \statusWarn{} -- upload failed, but earlier plan continued
\end{table}

To comply with radio frequency regulations, no real data other than the satellite's telemetry was transmitted using HSL and ISL during the experiments.
This was realized on the satellite by configuring the radio module to not use the power amplifier (PA) connected to the antenna.
To compensate for the energy consumption of the PA, the GPS payload was simultaneously activated during HSL and ISL tasks.

The scenario can be easily chained up to arrive at larger satellite constellations with non-trivial communication opportunities.
Since future \gomspace missions will exploit ISLs for continuous networked operation, interdependencies in data flow models are appealing extensions of the present work, where scalability needs to be mastered in terms of satellite fleet and orbital parameter diversity.\cite{DBLP:journals/ijscn/FraireGHNBB21}

\subsection{Evaluation}

The entire time of exclusive experimental access to \gomx{4} spanned a period of 758 consecutive hours in which \gomx{4A} was operated by the \powverHOOP toolchain.
In the following, we are focusing on two days from our experimental case study.
Starting on May 10, 2021, at 00:57, five Aalborg ground station passes were targeted for the consecutive upload of flight plans.
To obtain a reliable uplink, only passes with a maximum elevation higher than \ang{25} were considered.
This means that five schedules were made available in total, each having a scheduling horizon of 24 hours.
The precise timings when each schedule became valid are listed in \autoref{tab:experiment-summary}.

\paragraph{\LEOPOWVER in Action.}
Out of the five flight plans, four were uploaded successfully to the satellite.
The last plan suffered from transmission issues on the ground segment and could not effectuate on the satellite.
Even with the best precautions, such radio communication failures can occur anytime in ground-to-satellite communication.
This stresses the importance to compute schedules that cover well beyond the next ground station passes.
In fact, this failure had little impact on the overall mission since the previously valid schedule simply continued to execute.
Although this may in general lead to a substantial drift between the predicted battery behavior and real battery voltage, the Deep Battery model takes care of that and corrects the predictions over time when new telemetry measurements arrive.

\begin{sidewaysfigure*}
    \centering%
    \includegraphics{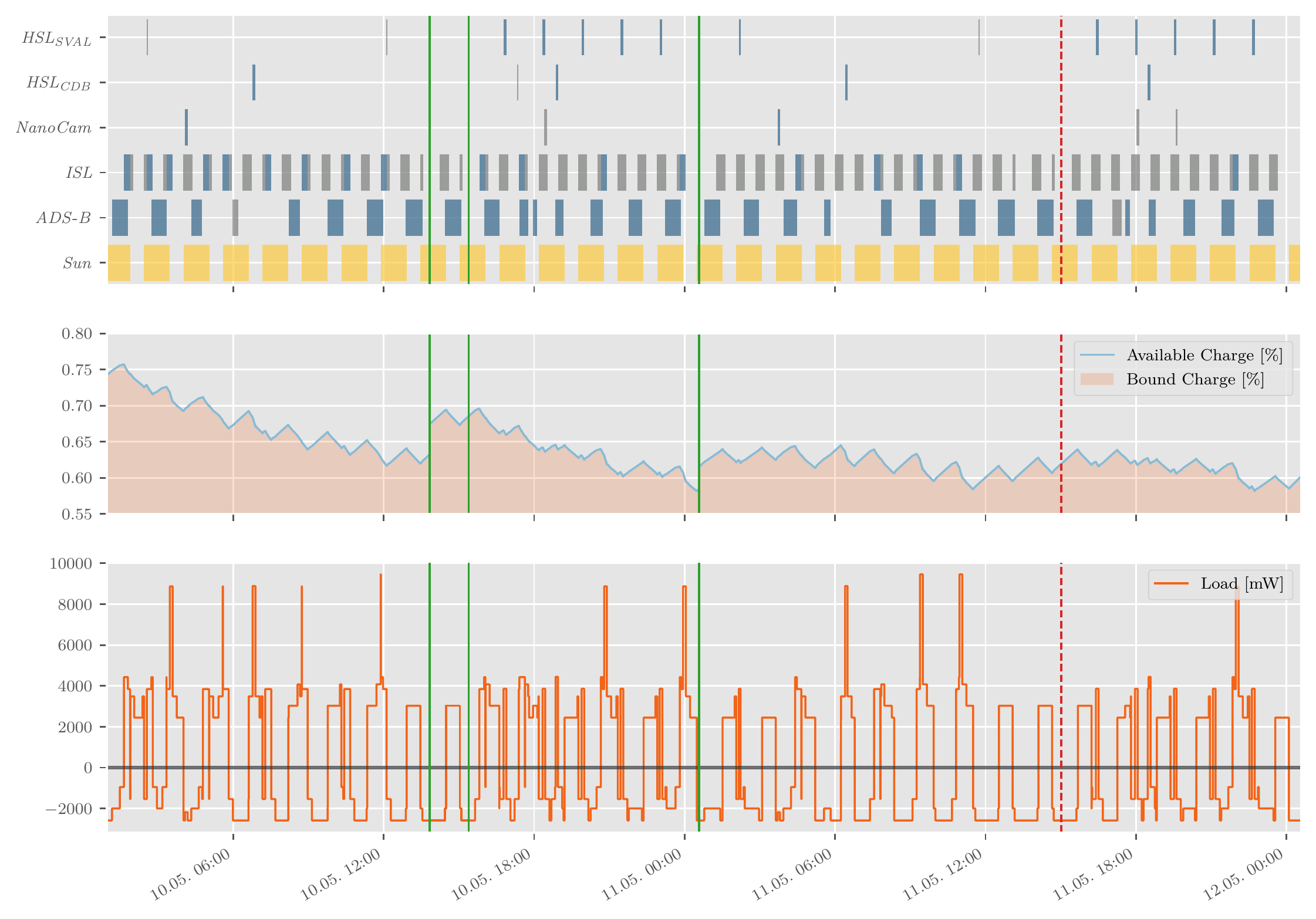}
    \caption{Tasks and predicted battery behavior of the receding-horizon schedules.}
    \label{fig:superimposed-schedules}
\end{sidewaysfigure*}

A temporal overview of the full experiment horizon with the four uploaded flight plans is shown in \autoref{fig:overview-four-plans}.
Each time the satellite passes over the ground station and a fresh and improved plan is uploaded to the satellite, the remaining parts of the previous plan are overwritten.
\autoref{fig:superimposed-schedules} illustrates this behavior for the whole experiment.
At the top, all possible payload utilization windows and sunlight exposure episodes are shown.
Access windows that were actually scheduled by our tool are marked in blue, or gray if skipped.
The three green lines mark the positions where new plans were uploaded (and the dashed red line where the upload failed).

\begin{figure*}
    \centering%
    \includegraphics{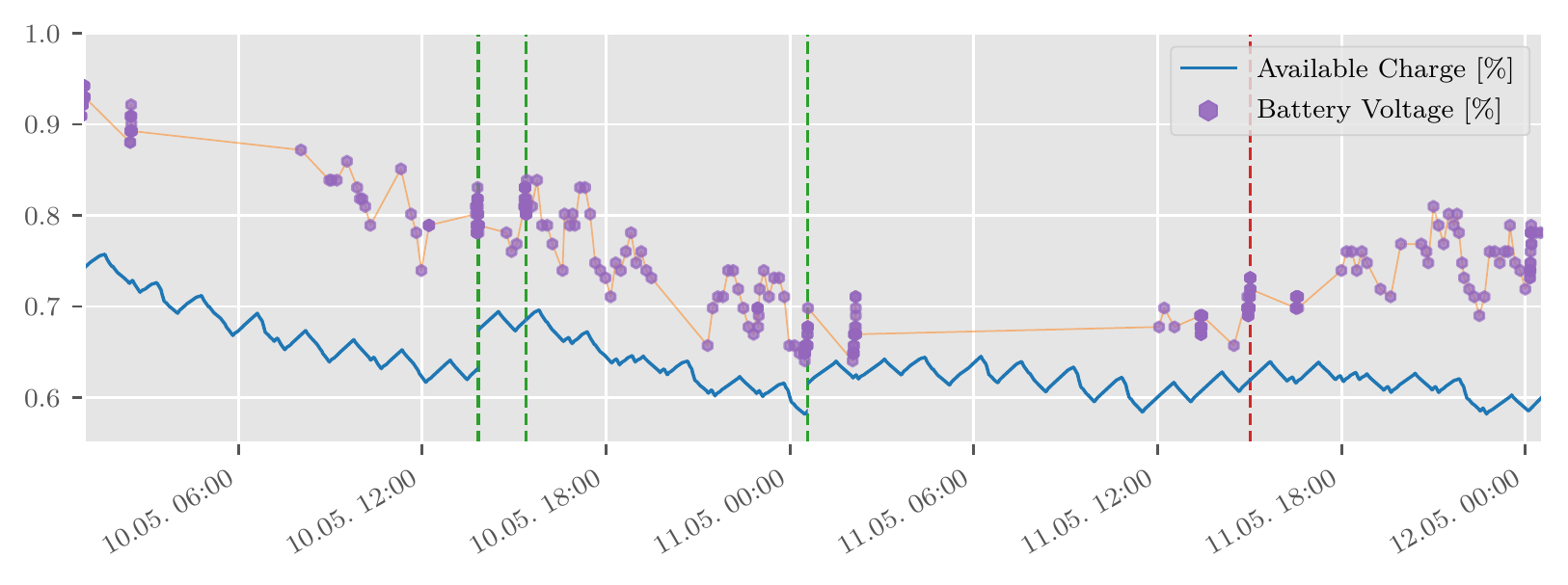}
    \caption{\gomx{4A} battery voltage telemetry during the experiment.}%
    \label{fig:telemetry-voltage}%
\end{figure*}

\paragraph{Predictions vs.\@ Reality.}
The middle and bottom of the figure show the predicted evolution of the battery with respect to the SoC of the \kibam model and the load profile that is applied to the battery.
According to telemetry measurements, the initial SoC was assumed to be around $\SI{75}{\percent}$, and the model predicted the SoC to decrease to $\SI{63}{\percent}$ at the moment the second flight plan should take over.
However, this was not the case either because the true initial SoC was actually higher than expected or because the model was too pessimistic concerning the payload energy utilization.
As a result, the initial SoC for the second plan was corrected upwards by the Deep Battery model by around $\SI{5}{\percent}$, allowing the second flight plan to perform more activities, hence increasing the satellite's productivity.
For the next pass 90 minutes later, a third flight plan was computed.
This time, the Deep Battery model induced no correction in either direction, meaning that the plan used the predicted model estimates unaltered to derive the initial SoC.
Similar to the second plan, the initial SoC of the fourth plan was again corrected upwards by the Deep Battery model.

The battery voltage measurements that were downlinked from the satellite and used in the Deep Battery model updates are shown in \autoref{fig:telemetry-voltage}.
The blue line again represents the evolution of the available charge of the KiBaM model.
The individual telemetry data points are depicted as purple hexagons, connected by an orange line that linearly interpolates the given discrete set of measurements recorded.
The battery pack on board \gomx{4A} has a maximal voltage of $\SI{16.2}{\V}$.
The lowest allowable operational limit for the battery is $\SI{14.8}{\V}$.
If the voltage falls below this nominal voltage limit, a safe-mode is triggered on the satellite that disables nonessential payloads until the battery has recovered.
According to the engineers, a battery pack with $\SI{14.8}{\V}$ contains roughly $\SI{55}{\percent}$ of energy.
Therefore, in the plot, a voltage of $\SI{14.8}{\V}$ is aligned with $\SI{55}{\percent}$.

\begin{sidewaysfigure*}
    \centering%
    \includegraphics{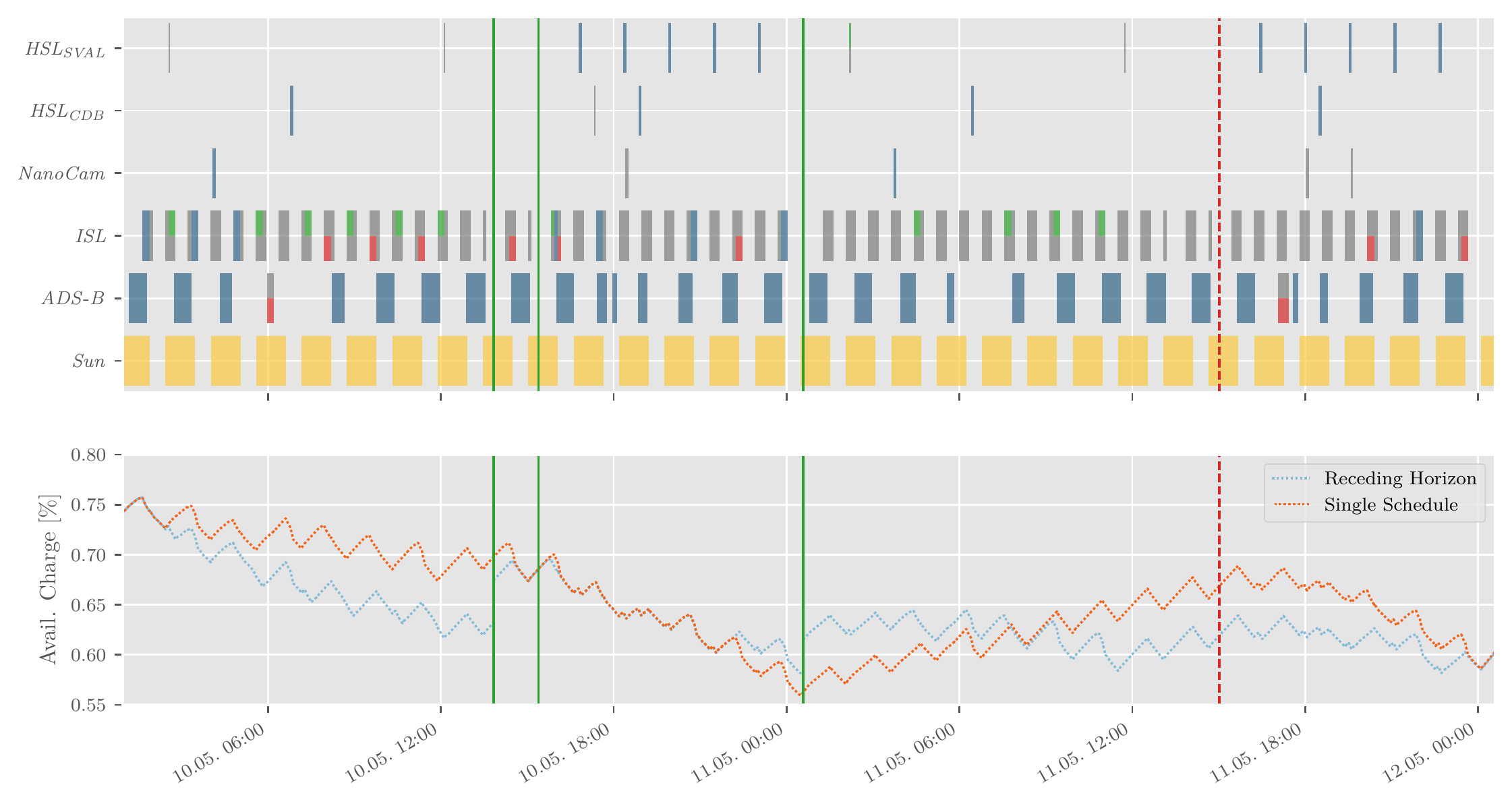}
    \caption{Comparison of the receding-horizon schedules with a monolithic schedule covering the entire experiment horizon.}%
    \label{fig:schedule-48h}%
\end{sidewaysfigure*}

\paragraph*{Effect of Receding-Horizon.}
For the purpose of analyzing the overall effect of the receding-horizon approach, a monolithic scheduling alternative was computed, using the same initial scenario setting as the real experiment but long enough to cover the entire experiment horizon in one shot, thus without considering any feedback from telemetry received during the experiment.
The receding-horizon schedules and the monolithic schedule are superimposed in \autoref{fig:schedule-48h}.
The payload utilization windows in the upper half of the figure are now split into two parts, where the top part belongs to the receding-horizon schedule, and the bottom part belongs to the single schedule.
Payload windows that are chosen by the tool to be executed only in the receding-horizon version but not in the single schedule are highlighted in green, while windows that are only scheduled in the monolithic schedule are colored red.

The computed schedules differ in many places, but if one looks at the total task counts being scheduled, the quantitative difference is that the receding-horizon schedule can sustain one more HSL task and six more ISL chunks than the single monolithic at the price of waiving two ADS-B opportunities.
Note, however, that from an energy budget perspective, HSL and ISL are much more power-demanding than ADS-B (around one order of magnitude).
These findings make sense since the Deep Battery model factually added energy to the system twice in the receding-horizon setting, based on the telemetry received.
This demonstrates the importance of automated model learning techniques whenever working with battery models based on some theoretical assumptions about the consumers.

\section{FUTURE DIRECTIONS}

This section provides a survey of lessons learned and further steps to be taken. 

\paragraph{HOOP and the \gomxx Series.}
The future direction as per \gomspace roadmaps are:

\begin{itemize}
    \item \textbf{\gomx{5}.}
          \gomspace is currently developing the next satellite in the \gomxx series, \mbox{\gomx{5}}.
          This mission, developed with the European Space Agency, is expected to launch in 2022 and will be fully operated through the HOOP platform, taking advantage of automation to optimize operations.
          \gomx{5} will demonstrate new capabilities in space, including increases in payload downlink volume, maneuverability for changes in orbit, and satellite position accuracy.
          The satellite will be equipped with advanced payloads and will carry out technology demonstration missions.
    \item \textbf{HOOP.}
          The development of HOOP is continuing, most recently under the MCOP Advanced project funded by the Luxembourg Space Agency and supported by the European Space Agency.
          This development includes advanced simulation capabilities, including enhanced flight dynamics tools and features to enable the management of larger constellations.
          HOOP will be deployed for an increasing number of commercial and scientific satellite missions, its automation capabilities enabling the operation of a growing number of smallsat constellations.
\end{itemize}

\paragraph{\LEOPOWVER.}
The following technical challenges are examples for the next steps that are planned to be tackled by the \LEOPOWVER ambitions.
They will be embedded into user-centric activities especially targeting start-ups and business customers.

\begin{itemize}
    \item \textbf{Data Flow Modeling.}
          The accurate commanding of space-to-ground and space-to-space communication subsystems depends on the information that is to be transferred via such links.
          The data volume to be transported is, in most cases, well known in advance as it is either generated periodically (telemetry) or triggered by commands sent in advance (science or mission data).
          In this context, well-known multi-commodity flow models and algorithms can be leveraged.\cite{DBLP:journals/adhoc/FraireMF16}
          Indeed, we are putting the technology in place for \LEOPOWVER to account for accurate data flow models to ensure more tight control of the utilization of power-hungry communication resources.\cite{DBLP:journals/tgcn/FraireNGHBB20}
    \item \textbf{Expression of Mission Objectives.}
          The current toolchain provides the mission operator with convenient options to define task importances and thus to guide the automated operations.
          However, complex constellation operations will require more powerful semantics to characterize specific mission goals.
          For example, tasks in constellations might be satellite-agnostic in the sense that acquisitions over a specific region can be executed by any of the satellites in the constellation. 
          Furthermore, task dependencies are a topic, \eg, if some data acquisition task is scheduled, then some downlink task should follow in the near future.
          The configuration of these constraints is possible already but can become more flexible, and indeed we are about to release an expressive language for tasking constellations.
          Algorithmically, these aspects can be framed as variants of the more general assignment problem, for which powerful algorithmic approaches are readily at hand at Saarland University to boost the capabilities of \LEOPOWVER.
    \item \textbf{Battery Degradation.}
          Batteries are known to degrade over time, meaning that they lose more and more storage capacity.
          While battery degradation cannot be avoided, being able to accurately predict and model it is crucial to increase the operational lifetime of the mission.
          The \LEOPOWVER Deep Battery model will be extended to learn on-the-fly when the battery behavior changes and thus predict the battery end-of-life performance.\cite{DBLP:conf/formats/WognsenHJHL15}
          This important phenomenon can then be considered during the scheduling phase, \eg, so that the affected satellite is scheduled to perform on nonessential tasks.
          When used in constellations of nodes with similar battery models, the learning could further profit from multiple data points.
\end{itemize}
In fact, these lists of action points appear a lot more separate than they actually are.
The experiments and the collaboration between the teams at \gomspace and Saarland University have been so successful that the technical and algorithmic progress made on the \LEOPOWVER side likely will go hand-in-hand with the advancements planned for HOOP. 

\section{CONCLUSION}

Driven by recent breakthroughs in technology, the interest of the space community in LEO satellites has reached an exorbitant level.
Thousands of scheduled launches are planned for years to come, opening the door for a sharply increasing market.
The skyrocketing number of satellites in orbit calls for the need for automatic satellite operation and management solutions.

\gomspace and Saarland University have successfully cooperated to combine their respective expertises: \gomspace as an industrial market leader in the commercialization of nanosatellites and Saarland University as distinguished technology provider regarding ultra-effective energy-aware cost-optimal task scheduling.
We have presented \powverHOOP, a fully automated satellite constellation operation framework, exploiting optimal computing techniques such as dynamic programming and self-learning models.
The dual-satellite \gomx{4} mission by \gomspace was used to validate and demonstrate the approach in orbit, allowing our integrated software to be distinguished as \emph{orbit-proof}. The \LEOPOWVER software is unique in its features:
\begin{itemize}
    \item The Deep Battery model exploits the non-linear dynamics of the kinetic battery model and is self-adjusting continuously with the latest telemetry received from orbit.
    \item The scheduling is perpetuated by a receding-horizon strategy.
    \item Data transfer between different satellites in orbit is accounted for by providing support for the scheduling of inter-satellite links.
    \item The core algorithmic problem is framed as dynamic programming with antichain-based pruning.
    \item The resulting software framework is directly applied for the automated management and operation of the \gomx{4} LEO mission.
    \item The approach is carefully crafted to take into account the usability by the space engineers and robustness against failures of parts of the toolchain.
    \item An extensive in-orbit test campaign validates accuracy, efficiency, scalability, and robustness with respect to the operational requirements and constraints of LEO constellations.
    \item The software architecture is designed in such a way that it can flexibly be embedded into the entire range of applications in LEO (and beyond).
          Connecting it to HOOP has been very straightforward and successful.
\end{itemize}

Further improvements and enhancements are on the roadmap.
The resulting machine learning approach from \LEOPOWVER as well as the operation scope of HOOP are highly flexible and can be scaled to more complex networked constellation-class missions.
In this view, more elaborated data flow and battery models and more powerful mission semantics are being integrated into \LEOPOWVER and thus made available to operators of larger missions via the enhanced HOOP interface.

\subsection{Acknowledgments}
This research has received support by the ERC Advanced Grant 695614 (\href{https://powver.org}{\POWVER}), the ERC Proof of Concept Grant 966770 (\href{https://leopowver.space}{\LEOPOWVER}), and by the DFG Grant 389792660, as part of TRR 248 (\href{https://perspicuous-computing.science}{CPEC}).

\pdfbookmark[1]{References}{references}
\printbibliography

@article{DBLP:journals/adhoc/FraireMF16,
  author    = {Juan A. Fraire and Pablo G. Madoery and Jorge M. Finochietto},
  title     = {Traffic-aware contact plan design for disruption-tolerant space sensor networks},
  journal   = {Ad Hoc Networks},
  volume    = {47},
  pages     = {41--52},
  year      = {2016},
  _url      = {https://doi.org/10.1016/j.adhoc.2016.04.007},
  doi       = {10.1016/j.adhoc.2016.04.007},
  timestamp = {Fri, 30 Nov 2018 13:24:42 +0100},
  biburl    = {https://dblp.org/rec/journals/adhoc/FraireMF16.bib},
  bibsource = {dblp computer science bibliography, https://dblp.org}
}

@inproceedings{refaat2018high,
  title     = {High Accuracy Spacecraft Orbit Propagator Validation},
  author    = {Ahmed Refaat and Ahmed Badawy and Mahmoud Ashry and Adel Omar},
  booktitle = {Proceedings of the 18th International Conference on Applied Mechanics and Mechanical Engineering},
  volume    = {18},
  year      = {2018},
  doi       = {10.21608/AMME.2018.34732},
  _url      = {https://doi.org/10.21608/AMME.2018.34732},
  publisher = {Military Technical College}
}

@inproceedings{vallado2006revisiting,
  author    = {David A. Vallado and Paul Crawford and Richard Hujsak and T. S. Kelso},
  title     = {Revisiting Spacetrack Report \#3},
  booktitle = {AIAA/AAS Astrodynamics Specialist Conference and Exhibit},
  year      = {2006},
  _url      = {https://doi.org/10.2514/6.2006-6753},
  doi       = {10.2514/6.2006-6753}
}

@article{DBLP:journals/cm/FraireF15,
  author    = {Juan A. Fraire and Jorge M. Finochietto},
  title     = {Design challenges in contact plans for disruption-tolerant satellite networks},
  journal   = {{IEEE} Commun. Mag.},
  volume    = {53},
  number    = {5},
  pages     = {163--169},
  year      = {2015},
  _url      = {https://doi.org/10.1109/MCOM.2015.7105656},
  doi       = {10.1109/MCOM.2015.7105656},
  timestamp = {Tue, 25 Aug 2020 16:47:14 +0200},
  biburl    = {https://dblp.org/rec/journals/cm/FraireF15.bib},
  bibsource = {dblp computer science bibliography, https://dblp.org}
}

@inproceedings{DBLP:conf/formats/WognsenHJHL15,
  author    = {Erik Ramsgaard Wognsen and Boudewijn R. Haverkort and Marijn R. Jongerden and Ren{\'{e}} Rydhof Hansen and Kim Guldstrand Larsen},
  editor    = {Sriram Sankaranarayanan and Enrico Vicario},
  title     = {A Score Function for Optimizing the Cycle-Life of Battery-Powered Embedded Systems},
  booktitle = {Formal Modeling and Analysis of Timed Systems - 13th International
               Conference, {FORMATS} 2015, Madrid, Spain, September 2-4, 2015, Proceedings},
  series    = {Lecture Notes in Computer Science},
  volume    = {9268},
  pages     = {305--320},
  publisher = {Springer},
  year      = {2015},
  _url      = {https://doi.org/10.1007/978-3-319-22975-1\_20},
  doi       = {10.1007/978-3-319-22975-1_20},
  timestamp = {Sat, 19 Oct 2019 20:37:31 +0200},
  biburl    = {https://dblp.org/rec/conf/formats/WognsenHJHL15.bib},
  bibsource = {dblp computer science bibliography, https://dblp.org}
}

@article{DBLP:journals/tgcn/FraireNGHBB20,
  author    = {Juan A. Fraire and Gilles Nies and Carsten Gerstacker and Holger Hermanns and Kristian Bay and Morten Bisgaard},
  title     = {Battery-Aware Contact Plan Design for {LEO} Satellite Constellations: The Ulloriaq Case Study},
  journal   = {{IEEE} Trans. Green Commun. Netw.},
  volume    = {4},
  number    = {1},
  pages     = {236--245},
  year      = {2020},
  _url      = {https://doi.org/10.1109/TGCN.2019.2954166},
  doi       = {10.1109/TGCN.2019.2954166},
  timestamp = {Thu, 18 Jun 2020 22:04:09 +0200},
  biburl    = {https://dblp.org/rec/journals/tgcn/FraireNGHBB20.bib},
  bibsource = {dblp computer science bibliography, https://dblp.org}
}

@article{DBLP:journals/ijscn/FraireGHNBB21,
  author    = {Juan A. Fraire and Carsten Gerstacker and Holger Hermanns and Gilles Nies and Morten Bisgaard and Kristian Bay},
  title     = {On the scalability of battery-aware contact plan design for {LEO} satellite constellations},
  journal   = {Int. J. Satell. Commun. Netw.},
  volume    = {39},
  number    = {2},
  pages     = {193--204},
  year      = {2021},
  _url       = {https://doi.org/10.1002/sat.1374},
  doi       = {10.1002/sat.1374},
  timestamp = {Tue, 02 Mar 2021 11:26:05 +0100},
  biburl    = {https://dblp.org/rec/journals/ijscn/FraireGHNBB21.bib},
  bibsource = {dblp computer science bibliography, https://dblp.org}
}

@article{DBLP:journals/fac/BisgaardGHKNS19,
  author    = {Morten Bisgaard and
               David Gerhardt and
               Holger Hermanns and
               Jan Krč{\'a}l and
               Gilles Nies and
               Marvin Stenger},
  title     = {Battery-aware scheduling in low orbit: the GomX-3 case},
  journal   = {Formal Aspects Comput.},
  volume    = {31},
  number    = {2},
  pages     = {261--285},
  year      = {2019},
  _url      = {https://doi.org/10.1007/s00165-018-0458-2},
  doi       = {10.1007/s00165-018-0458-2},
  timestamp = {Tue, 25 Aug 2020 16:46:02 +0200},
  biburl    = {https://dblp.org/rec/journals/fac/BisgaardGHKNS19.bib},
  bibsource = {dblp computer science bibliography, https://dblp.org}
}

@article{DBLP:journals/iee/JongerdenH09,
  author    = {Marijn R. Jongerden and Boudewijn R. Haver\-kort},
  title     = {Which battery model to use?},
  journal   = {{IET} Softw.},
  volume    = {3},
  number    = {6},
  pages     = {445--457},
  year      = {2009},
  _url      = {https://doi.org/10.1049/iet-sen.2009.0001},
  doi       = {10.1049/iet-sen.2009.0001},
  timestamp = {Fri, 22 May 2020 15:37:01 +0200},
  biburl    = {https://dblp.org/rec/journals/iee/JongerdenH09.bib},
  bibsource = {dblp computer science bibliography, https://dblp.org}
}

@article{DBLP:journals/lites/HermannsKN17,
  author    = {Holger Hermanns and Jan Krč{\'{a}}l and Gilles Nies},
  title     = {How Is Your Satellite Doing? Battery Kinetics with Recharging and Uncertainty},
  journal   = {Leibniz Trans. Embed. Syst.},
  volume    = {4},
  number    = {1},
  pages     = {04:1--04:28},
  year      = {2017},
  _url      = {https://doi.org/10.4230/LITES-v004-i001-a004},
  doi       = {10.4230/LITES-v004-i001-a004},
  timestamp = {Thu, 10 Sep 2020 14:36:56 +0200},
  biburl    = {https://dblp.org/rec/journals/lites/HermannsKN17.bib},
  bibsource = {dblp computer science bibliography, https://dblp.org}
}

@article{DBLP:journals/msom/ChandHS02,
  author    = {Suresh Chand and Vernon Ning Hsu and Suresh P. Sethi},
  title     = {Forecast, Solution, and Rolling Horizons in Operations Management Problems: {A} Classified Bibliography},
  journal   = {Manuf. Serv. Oper. Manag.},
  volume    = {4},
  number    = {1},
  pages     = {25--43},
  year      = {2002},
  _url      = {https://doi.org/10.1287/msom.4.1.25.287},
  doi       = {10.1287/msom.4.1.25.287},
  timestamp = {Thu, 01 Oct 2020 10:17:34 +0200},
  biburl    = {https://dblp.org/rec/journals/msom/ChandHS02.bib},
  bibsource = {dblp computer science bibliography, https://dblp.org}
}

@article{DBLP:journals/tcad/StockFMHBC20,
  author    = {Gregory Stock and Juan A. Fraire and Tobias Mömke and Holger Hermanns and Fakhri Babayev and Eduardo Cruz},
  title     = {Managing Fleets of {LEO} Satellites: Nonlinear, Optimal, Efficient, Scalable, Usable, and Robust},
  journal   = {{IEEE} Trans. Comput. Aided Des. Integr. Circuits Syst.},
  volume    = {39},
  number    = {11},
  pages     = {3762--3773},
  year      = {2020},
  _url      = {https://doi.org/10.1109/TCAD.2020.3012751},
  doi       = {10.1109/TCAD.2020.3012751},
  timestamp = {Thu, 17 Dec 2020 18:29:36 +0100},
  biburl    = {https://dblp.org/rec/journals/tcad/StockFMHBC20.bib},
  bibsource = {dblp computer science bibliography, https://dblp.org}
}

@inproceedings{conf/smallsat/LeonKW18,
  author    = {Laura León Pérez and Per Koch and Roger Walker},
  title     = {GOMX-4 -- The Twin European Mission for IOD Purposes},
  booktitle = {Proceedings of the AIAA/USU Conference on Small Satellites},
  year      = {2018},
  number    = {SSC18-VII-07},
  series    = {Science / Mission Payloads,},
  url       = {https://digitalcommons.usu.edu/smallsat/2018/all2018/296/}
}

@inproceedings{conf/smallsat/GoldbergKRHTP19,
  author    = {Hannah R. Goldberg and Özgür Karatekin and Birgit Ritter and Alain Herique and Paolo Tortora and Claudiu Prioroc and Borja Garcia Gutierrez and Paolo Martino and Ian Carnelli},
  title     = {The Juventas CubeSat in Support of ESA's Hera Mission to the Asteroid Didymos},
  booktitle = {Proceedings of the AIAA/USU Conference on Small Satellites},
  year      = {2019},
  number    = {SSC19-WKIV-05},
  series    = {Instruments / Science,},
  url       = {https://digitalcommons.usu.edu/smallsat/2019/all2019/73/}
}

@inproceedings{conf/smallsat/StockFHCII21,
  author    = {Stock, Gregory and Fraire, Juan A. and Hermanns, Holger and Cruz, Eduardo and Isaacs, Alastair and Imbrosh, Zhana},
  title     = {On the Automation, Optimization, and In-Orbit Validation of Intelligent Satellite Constellation Operations},
  booktitle = {Proceedings of the AIAA/USU Conference on Small Satellites},
  year      = {2021},
  number    = {SSC21-V-05},
  series    = {Ground Systems,},
  url       = {https://digitalcommons.usu.edu/smallsat/2021/all2021/168/}
}

@article{journals/actaastro/NiesSKHBG18,
  author  = {Gilles Nies and Marvin Stenger and Jan Krč{\'a}l and Holger Hermanns and Morten Bisgaard and David Gerhardt and Boudewijn Haverkort and Marijn Jongerden and Kim G. Larsen and Erik R. Wognsen},
  title   = {Mastering operational limitations of {LEO} satellites -- {The} {GomX-3} approach},
  journal = {Acta Astronautica},
  volume  = {151},
  pages   = {726--735},
  year    = {2018},
  issn    = {0094-5765},
  doi     = {10.1016/j.actaastro.2018.04.040},
  _url    = {https://doi.org/10.1016/j.actaastro.2018.04.040}
}

@book{Bellman57,
  author    = {Richard E. Bellman},
  title     = {Dynamic Programming},
  publisher = {Princeton University Press},
  edition   = {1},
  year      = {1957}
}

@misc{GMAT,
  author  = {NASA Goddard Space Flight Center},
  title   = {{General Mission Analysis Tool (GMAT)}},
  note    = {Version R2020a},
  url     = {http://gmatcentral.org/},
  urldate = {2021-05-26}
}

@misc{OREKIT,
  author  = {CS Group},
  title   = {Orekit},
  note    = {Version 10.3},
  url     = {https://www.orekit.org/},
  urldate = {2021-05-18}
}

@misc{STK,
  author  = {{Analytical Graphics, Inc.}},
  title   = {{Systems Tool Kit (STK)}},
  note    = {Version 12.2},
  url     = {https://www.agi.com/products/stk/},
  urldate = {2021-05-18}
}

\end{document}